# Optimization of Bloch-Siegert B1 Mapping Sequence for Maximum Signal to Noise

M. Mehdi Khalighi, Doug Kelley, Jason H. Su, Brian K. Rutt and Adam B. Kerr

*Abstract*— Adiabatic Bloch-Siegert $B_1^+$ mapping method addresses the long TE and high RF power deposition problems of conventional Bloch-Siegert $B_1^+$ mapping by introducing short frequency-swept ABS pulses with maximum sensitivity. Here, it is shown how maximum signal to noise ratio can be achieved in adiabatic Bloch-Siegert $B_1^+$ mapping. Signal to noise ratio of $B_1^+$ maps is maximized by optimizing the adiabatic pulse parameters such as width, amplitude and shape of the Bloch-Siegert pulse within a specified scan time and under approved SAR guidelines. Equations for optimized Bloch-Siegert pulse parameters are derived, which are dependent on the base pulse sequence used for $B_1^+$ mapping as well as tissue properties and transmit coil configuration. It is shown that by this optimization it is more efficient to increase TR rather than using the averaging method to increase signal to noise ratio.

*Index Terms*— $B_1^+$ mapping, flip angle, Bloch-Siegert shift, signal to noise ratio, parallel transmit, adiabatic pulse

## I. INTRODUCTION

ONE of the major challenges in high field magnetic resonance is radio frequency (RF) shading due to the shortening of the RF wavelength and dielectric effect. One way to overcome this problem is Parallel Transmit (pTx) which is done either by optimizing the phase and amplitude of all Tx channels (i.e. RF Shimming) or by designing pTx RF pulses [1, 2]. For both cases, the first step is acquiring the RF magnetic field or $B_1^+$ map generated by each channel, which can become a long calibration scan. Bloch-Siegert (B-S) $B_1^+$ mapping method [3] is an accurate and efficient phase-based method that is largely insensitive to tissue properties such as T1 and T2, but it suffers from high RF deposition (SAR) and long TE, which results in long scan times. The Adiabatic Bloch-Siegert (ABS) $B_1^+$ mapping method [4] maximizes the Angle to Noise ratio (ANR) of the B-S method by using an adiabatic RF pulse as the B-S pulse which maximizes the

This project was supported by GE HealthCare. This work involved human subjects in its research. Approval of all ethical and experimental procedures and protocols was granted by Stanford's institutional review board (IRB).

Mehdi Khalighi is with Radiology Department, Stanford University (email: mkhalighi@stanford.edu).
Doug Kelley is with Radiology Department, Stanford University (email: mkhalighi@stanford.edu).
Jason H. Su is with Electrical Engineering Department, Stanford University (email: research@whiterabbit.ai).
Brian K. Rutt is with Radiology Department, Stanford University (email: brutt@stanford.edu).
Adam B Kerr is with Electrical Engineering Department, Stanford University (email: akerr@stanford.edu).

sensitivity and allows for shorter scan times.

In practice, the ABS pulse width and amplitude are selected to achieve high ANR $B_1^+$ maps. This will increase the RF deposition within repetition time (TR) and therefore increases the minimum TR due to SAR constraints. In other words, minimum TR and hence the scan time will be a function of ANR which in turn is a function of ABS pulse width and amplitude as well as tissue properties and pulse sequence. Here, we investigate how ABS pulse parameters affect scan time and ANR and find the best set of parameters to achieve the shortest scan time while respecting SAR limits.

## II. THEORY

Let's assume $S_i$ is the image magnitude for the *i*-th coil of a phased array receive coil with $N_c$ receive channels. Also assume that each channel has a Gaussian noise with standard deviation of σ which is uncorrelated with the other channels. If we assume $S_i \gg \sigma$, then the phase image noise has a Gaussian distribution with standard deviation of the *i*-th channel ($\sigma_{\phi i}$) is given by [5]:

$$\sigma_{\phi i} = \frac{\sigma}{S_i} \quad (1)$$

The B-S phase shift is calculated by subtracting the phases of two images acquired with B-S pulse at $\pm\Delta\omega_{RF}$ off-resonance frequency [3]. Assuming that these two images are statistically independent, we can write:

$$\phi_{BS_i} = \phi_{1_i} - \phi_{2_i} \Rightarrow \sigma_{\phi_{BS_i}} = \frac{\sqrt{2}\sigma}{S_i} \quad (2)$$

where *i* is the number of channel, $\phi_{1i}$ is the phase of image acquired with B-S pulse at $+\Delta\omega_{RF}$ and $\phi_{2i}$ is the phase of image acquired with B-S pulse at $-\Delta\omega_{RF}$. Now using the minimum variance estimator, the B-S phase shift is given by:

$$\phi_{BS} = \frac{\sum_{i=1}^{N_c} S_i^2 \phi_{BS_i}}{\sum_{i=1}^{N_c} S_i^2} \quad (3)$$

Therefore, the B-S phase variance is given by:

$$\sigma_{\phi_{BS}}^2 = \frac{2N_c\sigma^2}{\sum_{i=1}^{N_c} S_i^2} \quad (4)$$

Assuming the B-S pulse sequence is a spoiled Gradient Echo sequence [3] with an excitation flip angle of $\alpha$, therefore:

$$S_i = M_0 \frac{E_2^*(1-E_1)\sin\alpha}{1-\cos\alpha E_1}|R_i| \quad (5)$$

where $M_0$ is the initial magnetization, $R_i$ is the receive sensitivity of the $i$-th coil, and $E_1$ and $E_2^*$ are defined as:

$$E_1 = \exp\left(-\frac{T_1}{TR}\right) \quad E_2^* = \exp\left(-\frac{T_2^*}{TE}\right) \quad (6)$$

Let's define $SNR_i$ as the maximum intrinsic signal to noise ratio independent of relaxation effects:

$$SNR_i = \frac{M_0}{\sigma}\sqrt{\frac{1}{N_c}\sum_{i=1}^{N_c}|R_i|^2} \quad (7)$$

By combining equations 4, 5 and 7 we will get:

$$\sigma_{\phi_{BS}} = \frac{1-E_1\cos\alpha}{E_2^*(1-E_1)\sin\alpha}\frac{\sqrt{2}}{SNR_i} \quad (8)$$

To maximize ANR, we set the excitation flip angle to the Ernst angle:

$$\cos\alpha = E_1 \Rightarrow \sigma_{\phi_{BS}} = \sqrt{2\frac{1+E_1}{1-E_1}}\frac{1}{E_2^*SNR_i} \quad (9)$$

Considering the relationship between peak $B_1$ ($B_{1p}$) and B-S phase shift [3]:

$$\phi_{BS} = K_{BS}B_{1p}^2 \Rightarrow \sigma_{\phi_{BS}} = 2B_{1p}K_{BS}\sigma_{B_{1p}} \quad (10)$$

where $K_{BS}$ is a constant factor which depends on the pulse shape and off-resonance frequency. Using (9) and (10) we can show:

$$ANR_{BS} = \frac{B_{1p}}{\sigma_{B_{1p}}} = \sqrt{2\frac{1+E_1}{1-E_1}}E_2^*B_{1p}^2 K_{BS}SNR_i \quad (11)$$

To achieve the shortest scan times, ANR given in (11) should be maximized. This is done through optimizing the base sequence parameters to gain maximum signal to noise ratio (e.g. setting the excitation flip angle to Ernst angle for spoiled Gradient Echo sequence), picking a receive coil with highest intrinsic $SNR_i$ (e.g. phased array coils instead of Body coil), maximizing image $SNR_i$ by using all channels for excitation [6], maximizing peak $B_1$ within SAR limits and maximizing $K_{BS}$ by designing a highly sensitive off-resonance B-S pulse i.e. ABS pulses.

As it has been shown in [4], the ABS pulse that produces maximum B-S phase shift with a given pulse width ($T$) and peak amplitude ($B_{1p}$) is defined by:

$$B_1(t) = B_{1p}\exp(i\Delta\omega_{RF}(t)t)\prod\left(\frac{t-T/2}{T}\right) \quad (12)$$

$$\Delta\omega_{RF}(t) = \begin{cases}\gamma B_{1p}\cot(\psi) & \text{if } 0 < t < T/2 \\ \gamma B_{1p}\cot(2\psi_0 - \psi) & \text{if } T/2 < t < T\end{cases} \quad (13)$$

$$\psi(t) = \cos^{-1}\left(1 - \frac{\gamma B_{1p}}{K}t\right) \quad (14)$$

$$\psi_0 = \psi(t)\Big|_{t=T/2} = \cos^{-1}\left(1 - \frac{\gamma B_{1p}T}{2K}\right) \quad (15)$$

where $\gamma$ is the gyromagnetic ratio and $K$ (adiabatic factor) is a design parameter which determines the in-band excitation tolerance. In a special case where there is no off resonance spins, the B-S shift [4] is given by:

$$\phi_{BS} = \int_0^T \gamma\sqrt{B_{1p}^2 + (\Delta\omega_{RF}(t)/\gamma)^2}\,dt - \int_0^T \gamma\Delta\omega_{RF}(t)\,dt$$
$$= 2K(\psi_0 - \sin\psi_0) \quad (16)$$

Assuming large $K$ values and relatively short pulses and using Taylor's expansion for (15) and (16), the B-S phase shift can be approximated by:

$$\psi_0 \approx \sqrt{\frac{\gamma B_{1p}T}{K}} \Rightarrow \phi_{BS} \approx \frac{(\gamma B_{1p}T)^{\frac{3}{2}}}{3K^{\frac{1}{2}}} \quad (17)$$

To gain maximum ANR, the ABS pulse amplitude must be set to maximum within SAR limits to maximize the B-S phase shift. This means that the energy under the ABS pulse ($B_{1p}^2 T$) is constant which depends on the pulse sequence and repetition time i.e. TR. Therefore:

$$B_{1p} \propto T^{-\frac{1}{2}} \Rightarrow \phi_{BS} \propto T^{\frac{3}{4}} \quad (18)$$

The B-S pulse efficiency, $\Gamma$, is defined as the B-S phase shift adjusted by the $SNR_i$ loss due to signal decay caused by the echo time (TE) increase that results from the insertion of the ABS pulse into the pulse sequence [7]. If the based ABS pulse sequence is a Gradient Echo sequence, then the signal decay due to increased TE is governed by $T_2^*$ decay:

$$\Gamma_{GE} = \phi_{BS}\exp(-T/T_2^*) = K_\Gamma T^{3/4}\exp(-T/T_2^*) \quad (19)$$

where $K_\Gamma$ is a constant. The optimum pulse width which gives the maximum efficiency for a gradient echo-based ABS

sequence is:

$$\frac{d\Gamma_{GE}}{dT} = 0 \Rightarrow T_{opt_{GE}} = \frac{3}{4}T_2^* \quad (20)$$

For a Spin Echo based ABS sequence, the echo time is increased by *2T* due to the insertion of two ABS pulses before and after the refocusing pulse and the signal decay is governed by $T_2$ decay. Therefore:

$$\Gamma_{SE} = \phi_{BS} exp(-2T/T_2) = K_\Gamma T^{3/4} exp(-2T/T_2) \quad (21)$$

$$\frac{d\Gamma_{SE}}{dT} = 0 \Rightarrow T_{opt_{SE}} = \frac{3}{8}T_2 \quad (22)$$

The optimum ABS pulse width is given by (20) and (22) for Gradient Echo and Spin Echo based sequences respectively and the ABS pulse amplitude is set to maximum within SAR limits, which is controlled by TR. To find the relationship between ANR and scan time for an optimum ABS pulse, let's consider a Gradient Echo based ABS $B_1^+$ mapping sequence and assume that the excitation pulse is small comparing to the ABS pulse so that most of the deposited RF energy is coming from the ABS pulse. Therefore, we can write:

$$SAR \propto B_{1p}^2 \Rightarrow B_{1p} \propto SAR^{\frac{1}{2}} \Rightarrow B_{1p} \propto TR^{\frac{1}{2}} \quad (23)$$

Combining this with (17) and we will get:

$$B_{1p} \propto TR^{\frac{1}{2}} \Rightarrow \phi_{BS} \propto TR^{\frac{3}{4}} \Rightarrow ANR \propto TR^{\frac{3}{4}} \Rightarrow ANR \propto (Scan\ time)^{\frac{3}{4}} \quad (24)$$

## III. METHODS

Phantom experiments were performed to measure the ABS method's ANR. The $B_1^+$ maps in a silicon oil phantom on a 7T scanner (GE-Healthcare, Waukesha WI) were measured with a Nova 32ch Head coil (Nova Medical, Wilmington, MA) and a conventional gradient echo-based ABS method using a 6ms B-S pulse with amplitude set between 1-6μT. The scan parameters were set as follows: TR = 34 ms, TE = 9.7 ms, FA=30 deg, matrix = 64×64, bandwidth ±32 kHz and slice thickness = 5 mm. Each experiment was repeated 20 times and B-S phase standard deviation and ANR maps were generated. B-S $B_1^+$ maps were compared using analytically designed ABS pulses (4) and in two different sequence conditions, always at constant scan time: a) minimum TR with increasing number of averages b) single average with increasing TR, with TR-optimized ABS pulse redesigned for each TR. In the first method the B-S phase shift maps were averaged together to generate higher ANR maps. In all cases, SAR was held constant. The relative ANR of the two methods was compared as relative scan time increased from 1 to 5, referenced against $ANR_{min}$, the ANR at $TR_{min}$ and single average.

A numerically optimized ABS pulse was designed [4] with -40 dB in-band attenuation assuming ±500 Hz on-resonance bandwidth. The ABS pulse was inserted into a spiral sequence [8] for B-S $B_1^+$ mapping. The spiral sequence parameters were selected as minimum TE, minimum TR, 2048 points, 2 arms, bandwidth ±83.3kHz, FOV 24 cm, slice thickness 5 mm, flip angle 30 deg and 25 contiguous slices. The sequence minimum TR and minimum TE were 850 ms and 9.9 ms respectively. Total scan time was 5 s for combined $B_1^+$ and B0 mapping. The sequence was used on a 7T GE scanner (GE Healthcare, Waukesha, WI) using a 32ch Nova Head coil (Nova Medical, Wilmington, MA). Based on the SAR limit, the peak amplitude (B1p) values for 27 different ABS pulses with various pulse widths T in the range 1-14 ms with 0.5 ms step were calculated for minimum TR. The B-S phase shift of each pulse was calculated using a Bloch simulation and the efficiency of each pulse was obtained from (19) for $T_{2^*}$ values over the range of [1, 30] ms. A subject was scanned and the $B_1^+$ map was acquired using 2, 4 and 6 ms ABS pulses. The scan was repeated 20 times for the central slice using each pulse and ANR maps were generated.

## IV. RESULTS

Fig. 1 shows the ANR maps of the ABS method in a silicon oil phantom with different ABS pulse amplitude. Higher ABS pulse amplitudes generate more B-S phase shift as is shown in the first row. The B-S phase standard deviation is shown in the 3rd row and as expected, is in the same range for all experiments (4). The last two rows show the ABS method's ANR and its mean value over the phantom and, as expected, the mean ANR increases quadratically with ABS pulse amplitude (11). Fig. 2-a shows the ABS pulse amplitude and phase modulation profiles along with the transverse magnetization frequency response. Fig. 2-b shows the B-S pulse efficiency plotted as a function of T and $T_{2^*}$ showing how the pulse efficiency decreases as $T_{2^*}$ decreases. The efficiency of longer ABS pulses decreases faster due to the longer echo time. The pulses that create the maximum efficiency for each $T_{2^*}$ value were picked from Fig. 2-b and those optimal pulse widths vs. $T_{2^*}$ are plotted in Figure 2-c. For comparison the optimum pulse width for the analytical design (20) is also plotted (blue line) and a good match with the numerically designed ABS pulse width vs. $T_{2^*}$ behavior is observed, with slope equal to 3/4. As $T_{2^*}$ is variable around the brain, 2, 4 and 6ms pulses were chosen for ANR comparison in head at 7T.

Fig. 3 compares the ANR obtained by these 3 pulses. It shows that the 6 ms ABS pulse is the most efficient of all 3 pulses under the same scan-time and SAR constraint. It also shows ANR is greater than 200 for most of the brain;

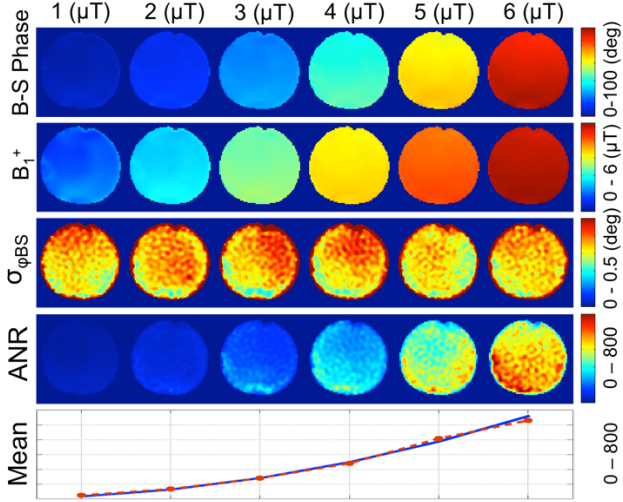

**Fig. 1**. Measuring ANR of ABS method using silicon oil at 7T with 1-6 µT ABS pulse amps. The ANR (red dots) increases quadratically with ABS pulse amplitude (blue line).

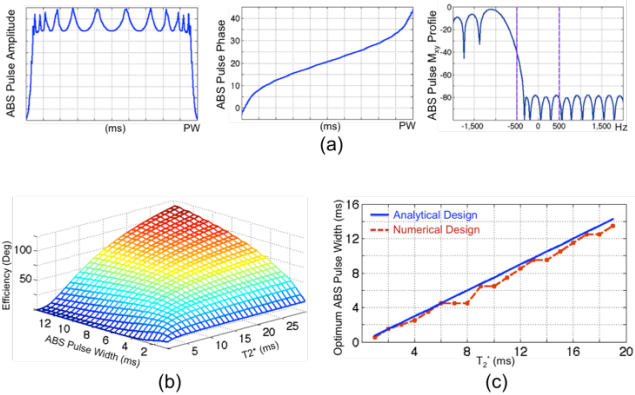

**Fig. 2.** (a) ABS pulse amplitude and phase along with transverse magnetization profile. (b) ABS pulse efficiency of numerically designed ABS pulses with different $T_{2^*}$ value. (c) Optimum ABS pulse width with different $T_{2^*}$ values using numerical and analytical methods.

however, it depends on the coil sensitivity, excitation flip angle, T1 and $T_{2^*}$ as it is shown by (11). Fig. 4 shows the whole brain $B_1^+$ maps using a 6 ms ABS pulse with minimum TR on a 7T GE scanner (GE Healthcare, Waukesha, WI). It shows that ABS method, which does not have any $T_1$ or $T_2$ dependency, generates whole brain high ANR $B_1^+$ maps using spiral readout in 5 s. Although the maps show sufficient ANR, by increasing TR higher ANR maps can be generated.

## V. DISCUSSION

The ANR equation for ABS method has been derived including effects of relaxation. Phantom and volunteer experiments have been performed to show the high ANR maps achieved by ABS. It was shown that instead of averaging, higher ANR maps can be achieved by increasing TR with optimization of ABS pulse design at the increased TR. In other words, for any given TR, there is an optimum ABS pulse amplitude, which uses all the SAR available within the sequence and increasing TR will increase the final ANR. In B-S $B_1^+$ mapping the imaging phase and $B_1^+$ encoding phase are separate and can be optimized individually [6]; however, this optimization is dependent on tissue properties and/or on $B_1^+$ uniformity. For instance, if the base sequence of B-S $B_1^+$ mapping is a Gradient Echo sequence then the optimal flip angle to get maximum ANR is the Ernst angle which depends on T1 at each voxel. On the other hand, flip angle is not uniform in high field MRI, and it varies from voxel to voxel, therefore the imaging part can be only optimized for a specific tissue at a specific location. ABS pulse optimization depends on $T_{2^*}$ or T2 depending on whether the B-S $B_1^+$ mapping is based on Gradient Echo or Spin Echo sequence, which depends on the magnetic field homogeneity and/or tissue properties.

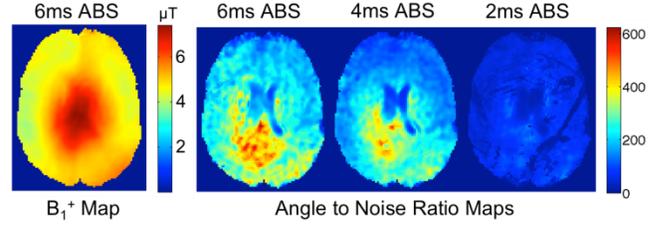

**Fig. 3.** ANR maps of brain $B_1^+$ map (left) acquired with ABS using 2, 4 and 6 ms pulse widths.

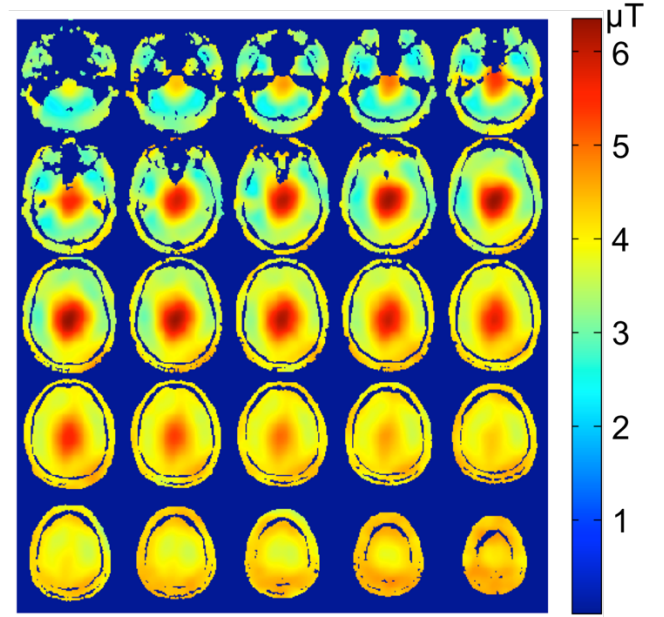

**Fig. 4.** $B_1^+$ maps at 7T by ABS method with spiral readout using a 6 ms ABS pulse with 6.3 µT amplitude. The total scan-time for whole brain $B_1^+$ and $B_0$ mapping is 5 s.

In summary, the optimum ANR will only be achieved over a (comparatively small) subset of voxels at 7T, although this effect may be minimal at 3T. What may matter more is that every voxel of interest exceeds a minimum ANR. The angle to noise ratio as defined in (11) concerns the statistical reproducibility of the $B_1^+$ maps, not necessarily their accuracy. Any bias [9] in the Bloch-Siegert measurement itself is not addressed by this analysis, nor is the effect of the signal-power-weighted combination of the phases from the different channels. It was shown that the optimum B-S pulse width for a Spin Echo based B-S $B_1^+$ mapping is longer than for a Gradient Echo based sequence, as signal decays faster in Gradient Echo based sequences. The maximum ANR achieved by optimizing B-S $B_1^+$ mapping sequence can be traded off for shorter scans. Shorter TR times require lower B-S pulse amplitudes within the SAR limit, which will decrease ANR. Picking shorter ABS pulse widths will have similar effect. To achieve very short scan times, Spiral readout was used; however, other accelerated readout trajectories such as Echo Planar readout [10] can be used. Other options for reaching shorter scan times are using parallel imaging techniques such as auto-calibrating methods [11] or SENSE [12]. However, due to the slow varying $B_1^+$ and therefore low resolution $B_1^+$ maps, the advantages of auto-calibrating parallel imaging methods are limited.

## VI. Conclusion

We have shown in this work that the Bloch-Siegert $B_1^+$ mapping sequence is optimized using an adiabatic off-resonance pulse and that the ABS pulse parameters i.e. pulse width and amplitude can be optimized given a specified scan time. It was shown that increasing scan time results in higher ANR compared to simple averaging and high-quality whole brain $B_1^+$ were acquired on a 7T scanner in only 5s.